\documentclass{article}

\usepackage{PRIMEarxiv}

\usepackage[utf8]{inputenc} 
\usepackage[T1]{fontenc}    
\PassOptionsToPackage{hyphens}{url}
\usepackage{hyperref}       
\usepackage{booktabs}       
\usepackage{amsfonts}       
\usepackage{nicefrac}       
\usepackage{microtype}      
\usepackage{lipsum}
\usepackage{fancyhdr}       
\usepackage{graphicx}       
\graphicspath{{media/}}     

\usepackage{tabularx}
\usepackage{booktabs}
\usepackage{ragged2e}

\usepackage{enumitem}

\usepackage{verbatim}       
\usepackage{xcolor}  
\usepackage{colortbl}
\usepackage{amsmath}

\usepackage{float}

\usepackage{multirow}

\usepackage{soul}

\definecolor{lightgray}{gray}{0.9}

\newcommand{\sunny}[1]{\textcolor{brown}{\textbf{Sunny:} #1}}

\newcommand{\harsha}[1]{\textcolor{violet}{\textbf{Harsha:} #1}}

\title{Achieving Responsible AI through ESG: \\
Insights and Recommendations from Industry Engagement}

\author{
   Harsha Perera, Sung Une Lee, Yue Liu, Boming Xia, Qinghua Lu, Liming Zhu \\
  CSIRO's Data61, Australia\\
   \AND
  Jessica Cairns, Moana Nottage \\
   Alphinity Investment Management, Australia \\
}

\begin{document}
\maketitle

\begin{abstract}
As Artificial Intelligence (AI) becomes integral to business operations, integrating Responsible AI (RAI) within Environmental, Social, and Governance (ESG) frameworks is essential for ethical and sustainable AI deployment. This study examines how leading companies align RAI with their ESG goals. Through interviews with 28 industry leaders, 
we identified a strong link between RAI and ESG practices. However, a significant gap exists between internal RAI policies and public disclosures, highlighting the need for greater board-level expertise, robust governance, and employee engagement. We provide key recommendations to strengthen RAI strategies, focusing on transparency, cross-functional collaboration, and seamless integration into existing ESG frameworks.
\end{abstract}


\section{Introduction} \label{sec:introduction}
Artificial Intelligence (AI), particularly exemplified by advancements in Generative AI such as Large Language Models (LLMs), has swiftly transformed various industries by driving significant efficiency gains and fostering innovation. However, the widespread integration of AI into critical domains and everyday life brings forth substantial risks, including privacy and copyright issues, potential biases, and security concerns \cite{yohsua2024international}. These challenges underscore the urgent need for robust governance frameworks to ensure that AI technologies are developed, deployed, and managed in a manner that is safe, transparent, and accountable. Consequently, Responsible AI (RAI) \cite{lu2023responsible} has emerged as a crucial foundation for safeguarding the ethical integrity and societal impact of AI.



In parallel, the Environment, Social, and Governance (ESG) framework has gained prominence as a tool for especially investors to evaluate non-financial risks and opportunities within companies. As AI becomes a central component of business operations and strategies, it is increasingly critical to assess its impact through the ESG lens. There is a growing consensus that AI’s ethical and societal implications cannot be effectively managed in isolation but must be integrated into broader corporate governance practices \cite{australiangovernment2024framework}. However, despite the alignment in goals, there remains a significant gap in how RAI principles are operationalized within existing ESG frameworks, leading to inconsistencies in governance and a potential lack of accountability.

This paper bridges this gap by advocating for the integration of RAI practices within the ESG framework, arguing that this synergy is essential for holistic and responsible AI governance. By embedding RAI within ESG, companies can ensure that their AI strategies not only comply with RAI standards but also contribute to sustainable and socially responsible business practices. To support this argument, we draw on insights from a collaborative research project between a leading research institute and an asset management firm in Australia \cite{ESGAIReport}. This collaboration involved in-depth interviews with 28 publicly listed companies, 
focusing on the practical challenges and opportunities in aligning RAI with ESG principles.

Our research reveals a persistent disconnect between internal RAI policies and their public disclosures, indicating a need for greater transparency and accountability. Companies often struggle to translate their RAI commitments into actionable strategies that align with their broader ESG objectives, which can undermine both their ethical stance and their attractiveness to socially conscious investors. This paper provides actionable recommendations, along with supporting tools and techniques, to bridge this gap.

The contributions of this paper are threefold: (1) it highlights the importance of integrating RAI within ESG frameworks to enhance corporate accountability; (2) it provides empirical evidence from leading companies on the current state of RAI implementation; and (3) it offers practical recommendations for companies seeking to strengthen their RAI strategies through ESG alignment. By doing so, this paper aims to contribute to the ongoing discourse on AI governance and provide valuable insights for both corporate leaders and investors who are navigating the complexities of AI safety and responsibility.

\section{Integrating RAI and ESG}

ESG frameworks are essential tools for managing non-financial risks such as environmental impact, social responsibility, and corporate governance. However, they often overlook the unique ethical and governance challenges posed by AI. Integrating RAI into ESG strategies is crucial, as it not only mitigates AI-related risks but also enhances sustainable and ethical governance, thereby aligning AI initiatives with broader corporate objectives.

The integration of RAI into ESG frameworks can be understood through three key dimensions:

\begin{enumerate}
    \item \textbf{AI as a Mitigator of ESG Risks:} AI technologies can play a significant role in reducing ESG-related risks. For example, AI can optimize resource use, improve energy efficiency, and strengthen governance practices to ensure compliance with regulatory standards. AI-driven analytics, for instance, can reduce environmental footprints by enabling more efficient operations.

    \item \textbf{AI as a Driver of Positive ESG Outcomes:} Beyond risk mitigation, AI has the potential to enhance ESG outcomes by driving operational efficiencies, improving customer experiences, and advancing sustainability initiatives. For example, AI-powered solutions in healthcare can increase accessibility to services while minimizing waste, thus contributing to both social and environmental goals.

    \item \textbf{RAI as an Integral ESG Concern:} The ethical challenges associated with AI—such as algorithmic bias, data privacy, and cybersecurity—are closely aligned with ESG principles. Addressing these challenges through RAI practices is essential for ensuring transparency, fairness, and accountability, which are fundamental to both governance and social responsibility.
\end{enumerate}

Despite the clear synergies between RAI and ESG, our research identifies a significant gap in how organizations integrate RAI into their ESG frameworks. This gap often manifests as a disconnect between internal RAI policies and external disclosures, highlighting the need for improved transparency and accountability. Furthermore, existing ESG reporting standards, such as those from the Global Reporting Initiative (GRI) and the Sustainability Accounting Standards Board (SASB), currently lack specific guidance on addressing AI-related risks, complicating efforts to align RAI with ESG strategies.

Although governments and industry bodies, including the EU AI Act \cite{euaiact} and the NIST AI Risk Management Framework \cite{NIST_AIRMF}, have begun to acknowledge the importance of integrating AI considerations within ESG frameworks, these initiatives are still in their early stages and do not yet provide comprehensive coverage. Therefore, it is imperative for companies to proactively embed RAI within their ESG frameworks to ensure that AI practices contribute positively to their overall ESG objectives, ultimately fostering sustainable and responsible business operations.

\section{Methodology}



This study is primarily based on qualitative data gathered through interviews with representatives from 28 companies spanning eight industry sectors, supplemented by desktop research on these companies. 
The interviews and research were focused on understanding AI policies, implementation strategies, governance, and stakeholder engagement, with a particular emphasis on RAI and ESG considerations. 



\textbf{Project Participants:} The project team consisted of members with diverse backgrounds and expertise, including RAI researchers, ESG experts/investors, an industrial AI expert, a project manager, and a design thinking practitioner. The project began with a hybrid workshop to establish clear research objectives. During this workshop, concurrent discussions led to the finalization of several strategic goals for the subsequent engagement phases, including the selection of sectors and companies to investigate.

\begin{table*}[tbp]
\footnotesize
\centering
\caption{The list of engaged companies and a summary of the interview results; Out of 28 engaged companies, 17 are based in Australia. The remaining companies are globally distributed across Europe, Asia, and America.}
\label{tab:demographics}
\rowcolors{2}{white}{lightgray}
\begin{tabular}
{p{0.05\textwidth}p{0.25\textwidth}p{0.05\textwidth}p{0.07\textwidth}p{0.45\textwidth}}
\toprule
\multicolumn{2}{l}{\textbf{Company}} & \textbf{Sector} & \textbf{Country} & \textbf{Interview result} \\
\midrule
\#1 & Mass media company. & CS & Australia & Limited AI strategy; early RAI principles; digital-driven governance; focus on premium content. \\ 
\#2 & Multinational investment holding company. & CS & Global & Strong RAI awareness; limited AI expertise; unclear strategy; high-risk exposure. \\ 
\#3 & Telecommunications holding company. & CS & Global & Strong tech expertise; advanced digital strategy; moderate AI risk awareness. \\  
\#4 & Diverse range of businesses including retail and insurance. & CD & Australia & Launching AI pilots; voluntary data ethics; improving GenAI policy. \\ 
\#5 & E-commerce platform operating company. & CD & Global & Advanced digital strategy; strong AI integration; working on AI policy; limited risk details. \\ 
\#6 & World-renowned luxury car manufacturer. & CD & Global & Minimal AI adoption; values human touch; exploring limited productivity gains. \\ 
\#7 & Oil and gas exploration and production company. & E & Australia & Early digitization; limited AI use; strong privacy focus. \\ 
\#8 & Petroleum exploration and production company. & E & Australia & Strong digital focus; no AI strategy; early RAI framework; active employee engagement. \\ 
\#9 & Multinational oil and gas company. & E & Global & Advanced RAI principles; large AI team; rolling out digital twins; strong focus on AI assurance. \\ 
\#10 & Leading insurance company. & F & Australia & Strong Board experience; AI-focused on customer tools; robust governance framework. \\ 
\#11 & Investment management company. & F & Australia & Early RAI development; GenAI focus; limited ESG awareness. \\ 
\#12 & Multinational banking and financial services company. & F & Australia & Strong risk management; integrated AI in tech; positive employee feedback on productivity. \\ 
\#13 & One of the largest banks in Australia. & F & Australia & Strong AI focus; tech-expert directors; internal RAI framework; strategic steps less clear. \\ 
\#14 & Multinational banking and financial services company. & F & Australia & Mature AI use; strong data ethics; focus on fraud and customer service; concerned about AI risks. \\ 
\#15 & Major stock exchange company. & F & Australia & Limited use of AI; rely on human capital; good tech risk processes in place. \\ 
\#16 & Investment banking and financial service company. & F & Australia & AI Board education, mature risk management, limited employee engagement, RAI procurement as a risk. \\ 
\#17 & Engineering professional services company. & I & Australia & Developing AI strategy; Board engagement; growing AI team; limited risk management. \\ 
\#18 & One of the world's largest toll-road operators. & I & Australia & Experienced governance; no specific AI strategy; bias management recognized; unclear risk management. \\ 
\#19 & Energy technology company. & I & Global & Developing AI framework; unclear strategy; large internal AI team. \\ 
\#20 & Leading development and management property company. & R & Australia & No AI board expertise; advanced digital strategy; developing GenAI; strong ESG focus. \\ 
\#21 & Leading steel producer. & M & Australia & Internal RAI framework; strong leadership engagement; reasonable ESG and risk management awareness. \\ 
\#22 & Major iron ore mining company. & M & Australia & Strong digital strategy; AI focus on innovation; ESG awareness; no specific AI risk management. \\ 
\#23 & Enterprise software company. & IT & Australia & Limited/no AI Board capability, AI committee, RAI framework, AI strategy; no AI opportunity/risk awareness.  \\ 
\#24 & Semiconductor and display production equipment company. & IT & Global & Long-time AI use; AI enhances productivity and significant benefits (e.g., 20\% emissions reduction). \\ 
\#25 & Electronic measurement solutions company. & IT & Global & AI committee; internal RAI principles; large R\&D team; risk focus on security. \\ 
\#28 & Global consulting, technology, and outsourcing service company. & IT & Global & Strong AI expertise; risk-based compliance; public RAI policy; clear implementation and client communications. \\ 
\#27 & One of the world's largest software makers. & IT & Global & 150 RAI Champions; adherence to AI rules; no specific reporting line; well-resourced; systemic risk focus; published principles with clear guardrails. \\ 
\#28 & Global provider of critical communication solutions. & IT & Global & No formal AI strategy; CEO support; early AI risk assessment; developing governance and guardrails. \\ 
\bottomrule
\multicolumn{5}{l}{(Sector)} \\
\rowcolor{lightgray} \multicolumn{5}{l}{CS: Communication Services, CD: Consumer Discretionary, E: Energy, F: Financials, I: Industrials, R: Real Estate,} \\
\multicolumn{5}{l}{M: Materials, IT: Information Technology} \\
\bottomrule
\end{tabular}
\end{table*}

\textbf{Selection of Sectors and Companies:}
A key objective was to gather data from a diverse range of sectors and leading global companies to gain insights into RAI and ESG practices. 
Leveraging our networks, we invited over 100 companies to participate in the study, with 28 companies accepting the invitation. 
It is noteworthy that those who declined were hesitant to disclose details about their AI practices, often citing the immaturity of their RAI implementation. 
The 28 companies are across eight sectors, including Financial, Information Technology, Energy, Consumer Discretionary, and Industrial (Table~\ref{tab:demographics}). 
As shown in the table, 
around 60\% of companies based in Australia, complemented by significant representation from the US, Europe, and Asia.







\textbf{Desk Research and Questionnaire:} Before conducting the interviews, we performed desk research to gather publicly available information on each company, such as data from their websites, annual reports, media coverage, and publications. Based on this research, we developed an initial questionnaire focusing on ESG and RAI practices. Recognizing the varying levels of AI maturity among the companies, we tailored the questionnaire to meet the specific needs of each company. Some example questions are listed below:

\begin{itemize}
    \item How is AI being integrated into this business? What are the main applications? How do you see this developing in 5 years?

    \item Does the business have an RAI policy, framework, or embed RAI in other operational policies? – What are the main components? Internal or external? How is this embedded from a governance perspective?

    \item Are the people with the Board and/or executive that have specific expertise in this area?  If not, is training a focus?

    \item What metrics does the company use to measure responsible AI, including progress?
\end{itemize}

\textbf{Interviews:}
The primary engagement with each participating company consisted of a 1-hour interview with 2-5 interviewees.
The interviews were conducted by 2 ESG experts/investors, with 1-3 researchers joining the meetings most of the time.
To protect sensitive information, the interviews were not recorded. Instead, detailed notes were taken in real-time. 
These notes were then electronically documented and securely stored.
Table \ref{tab:demographics} presents the demographics of interviewed companies including brief descriptions and a summary of the interview results.

\textbf{Data analysis and synthesis:}
The data analysis was conducted using a bespoke framework focused on four key pillars: Board oversight, RAI commitment, RAI implementation, and RAI metrics, as outlined in the recent ESG-AI framework~\cite{lee2024integrating}.
This approach facilitated a focused evaluation of RAI implementation across the organizations. The results were independently analyzed by two groups— RAI researchers and ESG experts/investors. We then collaborated to discuss discrepancies and ambiguities, ensuring a robust synthesis of research findings by leveraging the combined expertise of both groups. 
Several workshops were conducted to identify key insights emerging from the research findings. 
These workshops brought together the project team to collaboratively analyze the data. Through structured discussions and interactive sessions, we identified the most critical insights, refine our understanding of the challenges and opportunities in RAI implementation, and develop actionable recommendations for companies to enhance their ESG and RAI practices.

\section{Company Insights}



This section presents a summary of 10 key insights (IN1 - IN10) derived from an analysis of RAI practices, with a focus on identifying gaps in public disclosures, assessing AI implementation maturity, understanding employee engagement levels, and evaluating board capabilities.
These insights are organized systematically to align with the board-level oversight, RAI commitment, implementation, and metrics dimensions outlined in a recent ESG-AI framework \cite{lee2024integrating}, which provides a structured approach to integrating RAI within corporate governance.

\begin{itemize}
    \item \textbf{Board oversight}
    \begin{itemize}
        \item [IN1.] Strengthening Board and leadership capability in AI, technology and ethics
    \end{itemize}
    \item \textbf{RAI commitment}
        \begin{itemize}
        \item [IN2.] Only a small percentage of companies publicly disclose their RAI policies
        \item [IN3.] Most companies are investing in AI, but RAI policies and reporting are still being developed
        \item [IN4.] Data privacy is a key ESG issue, but other topics are still important and may be overlooked
        \item [IN5.] Companies are using different strategies for navigating and managing RAI risk, but supply chain management can be overlooked
    \end{itemize}
    \item \textbf{RAI implementation}
        \begin{itemize}
        \item [IN6.] Global equities are at the forefront of AI implementation
        \item [IN7.] A balanced view of threats and opportunities is needed to mitigate harm and leverage AI benefits
        \item [IN8.] Employee engagement is essential to deliver AI‑related opportunities 
        \item [IN9.] RAI governance is best embedded within existing systems and processes
    \end{itemize}
    \item \textbf{RAI metrics}
        \begin{itemize}
        \item [IN10.] Strong track record in ESG performance is an indicator of confidence for investors
    \end{itemize}
\end{itemize}

This approach covers critical dimensions to understand company’s commitment to RAI, including capability, maturity, and transparency.
By incorporating case studies, we aim to illustrate challenges and highlight best practices in RAI.

\subsection{Board Oversight}
Board oversight is primarily associated with board accountability and capability. This dimension ensures that AI governance is integrated into corporate structures, with accountability and competence required for effective management and oversight of RAI practices.

More specifically, navigating the complexities of AI deployment requires tech-savvy board members \textbf{[IN1]}. Companies must invest in training programs to raise AI awareness among their leadership. Proactive approaches, such as those adopted by \textit{MercadoLibre} and \textit{ANZ}, enhance board capabilities and ensure informed decision-making. Increasing AI expertise at the board level is vital for effective governance and strategic oversight of AI initiatives.

\subsection{RAI Commitment}
This dimension is particularly related to company's transparency and responsible disclosure, addressing concerns about public RAI policy and sensitive AI use cases. 
While public RAI policy promotes user awareness of impactful AI interactions \cite{stahl2023embedding}, sensitive use cases are regarded as high-risk AI which can cause significant issues to health and safety or fundamental rights of natural persons. 

Current trends indicate that only 10\% of interviewed companies publicly disclose their RAI policies, despite 40\% having internal policies \textbf{[IN2]}. This disparity highlights a need for increased transparency and accountability within the corporate sector. Greater public disclosure of RAI policies can enhance stakeholder trust and demonstrate a company's commitment to RAI practices. A notable example of transparency is \textit{Commonwealth Bank}, which has published its inaugural AI policy, setting a benchmark for others to follow.

However, there is a noticeable gap between RAI activities and their reflection in public reports \textbf{[IN3]}. Comprehensive reporting, including AI in risk statements and annual reports, is essential for transparency and stakeholder trust. Companies must bridge this gap by integrating AI-related disclosures into their broader ESG reporting frameworks to provide a clear and accurate representation of their RAI efforts.

There is a common consensus among the engaged companies that data privacy and cybersecurity are primary ESG concerns related to AI \textbf{[IN4]}. However, other critical issues like human rights and modern slavery receive limited attention in AI discussions. Companies like \textit{Transurban} prioritize data privacy and governance, setting an example for addressing these concerns. Broadening the focus to include these under explored topics ensures a comprehensive approach to ESG in AI implementations. 

To ensure transparency and privacy, managing AI risks should extend to supply chains and strategic partnerships \textbf{[IN5]}. Companies must ensure their procurement processes incorporate RAI considerations. Strategic partnerships (e.g., the \textit{Commonwealth Bank}’s collaboration with \textit{H2O.ai}) highlight the importance of addressing AI risks throughout the supply chain. This approach ensures that ethical standards and operational integrity are maintained across all business operations.

\subsection{RAI Implementation}
RAI implementation should ensure the responsible use of AI in daily operations. This includes key considerations and practices such as dedicated RAI responsibility, employee awareness, system integration, and AI incident management and reporting.

Through company engagement, we have identified that global leaders in AI implementation showcase extensive use cases and strategic integration, demonstrating how AI can be harnessed for competitive advantage and operational efficiency \textbf{[IN6]}. Companies such as \textit{Shell} and \textit{MercadoLibre} exemplify this trend with their robust AI strategies. In contrast, some regional companies (e.g., Australian companies, particularly in the banking sector) are beginning to explore AI opportunities but lag behind their global counterparts. This gap underscores the need for regional firms to accelerate their AI adoption to remain competitive.

Another key insight in implementing RAI is that a balanced approach to AI is crucial, as overly cautious strategies can stifle innovation \textbf{[IN7]}. Establishing clear AI guardrails allows companies to leverage AI's benefits while mitigating associated risks. \textit{Accenture}'s significant investment in RAI and its strategic toolkit exemplify how organizations can balance caution with innovation, ensuring that AI deployments are both safe and forward-looking.

Most engaged companies are aware of the importance of employee engagement for effective AI implementation \textbf{[IN8]}. For example, cross-functional collaboration involving both technical and non-technical staff is crucial. Fostering a culture of innovation and rapid adaptation is vital for AI success. \textit{Shell}'s AI community, with over 11,000 members, exemplifies how widespread engagement can drive innovation. Comprehensive AI training is also essential to build a risk-aware corporate culture, ensuring employees understand the implications and potential risks associated with AI technologies.

More importantly, integrating RAI governance within existing corporate frameworks is essential for comprehensive risk management and RAI implementation \textbf{[IN9]}. A cross-disciplinary approach involving diverse expertise ensures that potential risks are addressed holistically. Robust governance structures, as seen in companies like \textit{Microsoft} and \textit{Wesfarmers}, demonstrate the importance of embedding RAI principles into core operations. This integration safeguards ethical AI deployment and promotes consistent, responsible practices.

\subsection{RAI Metrics}
It is crucial for companies to identify RAI metrics, measure and manage RAI performance, and share these with stakeholders \cite{xia2024towards}. Disclosing these metrics and their performance increases transparency, enhances stakeholder awareness of RAI practices, and ultimately empowers investors to make more informed decisions.

Moreover, strong ESG performance correlates with increased confidence in managing AI, indicating that companies excelling in ESG are better positioned to handle AI-related challenges \textbf{[IN10]}. Examples like \textit{Keysight} and \textit{Mirvac} illustrate how robust ESG frameworks can inform RAI practices. This correlation underscores the importance of integrating ESG principles and key metrics into AI strategies to enhance overall corporate responsibility and sustainability.



\section{Recommendations}
This section provides five key recommendations for enhancing RAI practices. While not exhaustive, these suggestions offer a solid starting point for developing more robust frameworks. These recommendations are grounded in real-world insights from leading companies, reflecting their best practices and challenges. Moreover, RAI practices can be supported by the right tools and techniques (e.g., RAI patterns) at various levels such as the industry level, organization level and team level. We identified applicable patterns from the \textit{Responsible AI Pattern Catalogue} proposed in \cite{lu2024responsible} that companies can implement alongside each recommendation. 

\subsection{Enhance Board and Leadership AI Capabilities}
This recommendation supports \textit{Board oversight}, addressing the first insight [IN1] in the previous section.

Companies with knowledgeable top management are better equipped to make informed decisions about AI strategies, balancing risks and opportunities effectively \cite{macus2008board}. This demonstrates a proactive approach to responsible AI, which can boost investor confidence by showcasing a commitment to ethical leadership and innovation. 

To achieve this, companies should i) identify gaps in AI and technology knowledge among current top management, ii) recruit new executives with strong AI and technology backgrounds, iii) implement ongoing training programs for existing top management to stay updated on AI developments and ethical considerations, and iv) establish an RAI advisory group within the executive team to focus specifically on AI and related technologies.

 As in RAI patterns, Board capability can be enhanced by organization-level RAI patterns such as \textit{Leadership Commitment for RAI\footnote{\url{https://research.csiro.au/ss/science/projects/responsible-ai-pattern-catalogue/leadership-commitment-for-rai/}}} and \textit{RAI Training\footnote{\url{https://research.csiro.au/ss/science/projects/responsible-ai-pattern-catalogue/rai-training/}}}.
These patterns can be used to secure a more robust and expedited implementation of leadership in RAI within the organization.

\subsection{Increase Transparency in RAI Policies}
A company's \textit{RAI commitment} can be strengthened through enhanced transparency and the adoption of comprehensive RAI policies and practices. This can be achieved by primarily developing and disclosing RAI policies [IN2], incorporating holistic RAI principles and practices within these policies, and considering the entire supply chain [IN4 and IN5].

More importantly, companies should publicly disclose their RAI policies to build stakeholder trust and demonstrate a commitment to responsible AI practices \cite{zhu2022ai, stahl2023embedding}. Transparency in AI governance can enhance a company’s reputation, align with growing demands for corporate accountability, and encourage investor confidence through clear, responsible practices.

Accordingly, companies should i) develop clear and comprehensive RAI policies if they are not already in place, 
ii) publish these policies on the company’s website and include them in annual reports and ESG disclosures, iii) highlight key principles and guidelines that govern the ethical use of AI within your organization, iv) provide case studies or examples of how these policies are implemented in practice. 

\textit{RAI Law and Regulation\footnote{\url{https://research.csiro.au/ss/science/projects/responsible-ai-pattern-catalogue/ai-regulation/}}} and \textit{RAI Standards\footnote{\url{https://research.csiro.au/ss/science/projects/responsible-ai-pattern-catalogue/rai-standard/}}} patterns at the industry level should be considered to develop RAI policies aligned with relevant regulations and standards. 
Additionally, organization-level patterns such as \textit{Code for RAI\footnote{\url{https://research.csiro.au/ss/science/projects/responsible-ai-pattern-catalogue/code-for-rai/}}} provide overarching guidelines for establishing a robust RAI framework and policies.   

\subsection{Promote Employees' AI Literacy}
To advance in the market, a company needs to improve its maturity level in \textit{RAI implementation} [IN6]. This can begin with ensuring that all stakeholders (e.g., employees) have a solid understanding of AI, the opportunities and risks associated with its use, and strong competency [IN7].

Engaged and informed employees can identify AI opportunities and manage risks effectively \cite{hughes2019artificial}. This approach supports a strong risk management culture and drives innovation, showing the company values RAI development and is committed to fostering an environment of continuous learning and ethical improvement. 
To achieve this, we recommend fostering a culture of curiosity and engagement around AI through training and awareness programs for all employees, including non-technical staff.

Specifically, companies should i) develop AI literacy programs that explain the basics of AI, its applications, and ethical considerations, ii) encourage collaboration between technical and non-technical staff to generate AI-related ideas and solutions,
iii) create an internal AI community or forum where employees can share knowledge, discuss AI developments, and collaborate on projects, and iv) provide opportunities for hands-on learning and experimentation with AI tools and technologies.

\textit{RAI Risk Committee\footnote{\url{https://research.csiro.au/ss/science/projects/responsible-ai-pattern-catalogue/rai-risk-committee/}}} pattern can help to monitor, provide feedback and support organizational culture and RAI training. \textit{RAI Training} pattern can be a useful tool to plan and implement of RAI training for the employees, and ultimately contribute to sharpen the employees’ skills in developing or using AI systems responsibly.

\subsection{Foster Cross-functional Employee Engagement}
As most companies are well aware of the importance of employee engagement, \textit{RAI implementation} can be accelerated by transparent communications, knowledge-sharing, and cultural innovation [IN8]. At the same time, it is crucial that \textit{RAI implementation}, including AI governance and management practices, avoids becoming siloed [IN9].

Accordingly, companies should embed AI governance within existing systems and processes, involving representatives from various disciplines to oversee AI strategy and ensure its alignment with corporate values, including ESG principles. More importantly, cross-disciplinary governance ensures that AI decisions are ethically sound and align with the company’s broader business objectives, protecting social license, customer trust, and data privacy \cite{smith2019designing}.

Interviews with companies reveal that those with robust ESG governance are better equipped to adapt to RAI challenges and integration compared to those lacking in this area. This approach reassures investors that the company is taking a holistic and responsible approach to AI, leveraging ESG and design capabilities to foster trust and ensure sustainable business practices. Strong ESG practices position companies to effectively navigate AI challenges, reflecting a well-rounded and forward-thinking governance strategy.

Regarding this concern, companies should i) form a cross-disciplinary AI governance committee that includes representatives from IT, legal, compliance, HR, and other relevant departments,
ii) develop a charter for the committee, outlining its responsibilities and decision-making authority, explicitly incorporating ESG principles to guide ethical and sustainable AI practices, iii) ensure regular meetings to review AI projects, assess risks, and ensure alignment with corporate values and ESG objectives, and iv) create clear reporting lines to top management and senior leadership, ensuring that ESG considerations are integral to AI governance discussions and decisions.

To support these practices, organization-level patterns such as \textit{RAI Risk Committee}, \textit{Standardized Reporting\footnote{\url{https://research.csiro.au/ss/science/projects/responsible-ai-pattern-catalogue/standardized-reporting/}}} and \textit{Role-Level Accountability Contract\footnote{\url{https://research.csiro.au/ss/science/projects/responsible-ai-pattern-catalogue/role-level-accountability/}}} can facilitate clear processes and roles and responsibilities for stakeholders. 
In addition, \textit{Diverse Team\footnote{\url{https://research.csiro.au/ss/science/projects/responsible-ai-pattern-catalogue/diverse-team/}}} and \textit{Stakeholder Engagement\footnote{\url{https://research.csiro.au/ss/science/projects/responsible-ai-pattern-catalogue/stakeholder-engagement/}}} patterns help ensure that a wide range of perspectives is considered, fostering more inclusive and ethical AI decision-making processes at the team level.

\subsection{Implement Robust RAI Reporting and Target Setting}
Our insights reveal that while companies have dedicated commitment to RAI initiatives, they lack considerations for RAI reporting [IN3]. 
There is a significant gap between RAI activities and their reflection in public reports. 
Moreover, there is a need for \textit{RAI targets and metrics}, in alignment with ESG topics and performance, to be identified, managed, and disclosed to external stakeholders [IN10].

We recommend developing comprehensive RAI reporting mechanisms and setting clear AI targets. Publicly reporting on AI initiatives and their alignment with ESG goals is essential. This may include establishing clear AI guardrails to mitigate risks while fostering innovation, ensuring a balanced and forward-thinking AI strategy.

Consistent and transparent reporting on RAI activities enables investors and stakeholders to assess the company’s commitment to responsible AI use. In particular, rigorous monitoring and transparent reporting of serious incidents involving high-risk AI are essential. This approach serves as a vital mechanism to capture real-world failures, prevent their recurrence, and generate valuable insights that benefit all stakeholders~\cite{mcgregor2021preventing}. Furthermore, setting and publicly sharing targets drives progress and accountability, showcasing leadership in AI governance. This can enhance investor confidence by demonstrating a clear, responsible strategy for AI development and implementation.

To achieve these goals, a set of RAI practices should be implemented. Specifically, i) establish clear, measurable targets for AI initiatives that align with the company’s ESG goals,
ii) develop a framework for tracking and reporting progress toward these targets, iii) include RAI reporting in annual reports, ESG disclosures, and other public communications, and iv) regularly review and update targets and reporting frameworks to reflect advancements in AI and evolving ethical standards.

As supporting tools, \textit{RAI Risk Committee} pattern can be primarily used to set and monitor RAI goals and targets, as well as to provide feedback and guidance to the project team. \textit{Standardized Reporting} pattern also helps establish a transparent and responsible reporting culture and practices around AI system development and use.
At the team level, \textit{Continuous Documentation Using Templates\footnote{\url{https://research.csiro.au/ss/science/projects/responsible-ai-pattern-catalogue/continuous-documentation-using-templates/}}} pattern serves as a tool for recording activities, logs, and events associated with AI systems.

\begin{figure*}[htb]
    \centering
    \includegraphics[width=\textwidth]{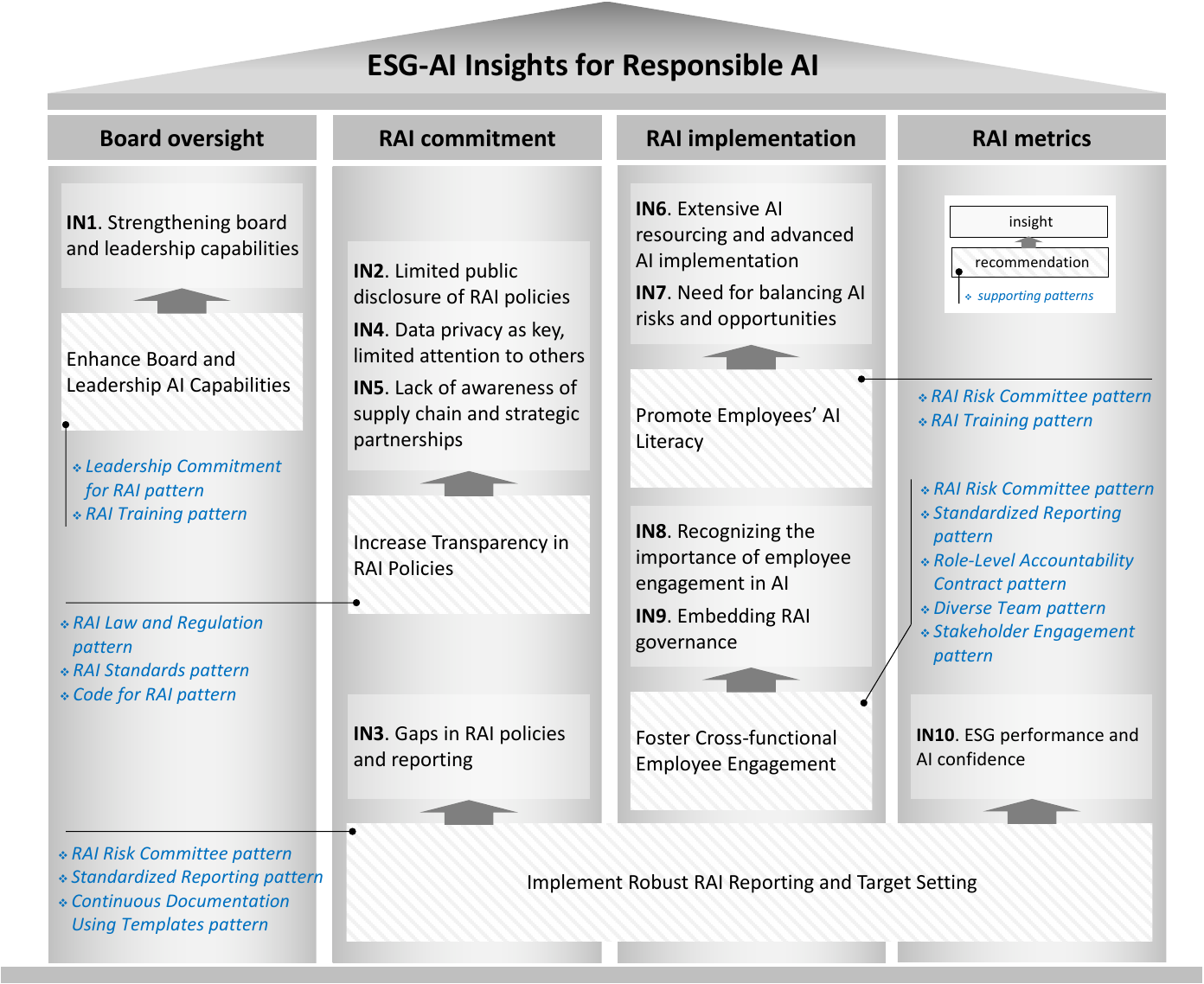}
    \caption{ESG-AI company insights and recommendations map; This includes a set of RAI patterns to support the recommendations.}
    \label{fig:Insights}
\end{figure*}

Figure \ref{fig:Insights} demonstrates the critical insights and recommendations under the four pillars: Board oversight, RAI commitment, RAI implementation, and RAI metrics, as well as supporting RAI patterns.

\newpage
\section{CONCLUSION}
The integration of RAI into existing ESG frameworks is crucial for aligning ethical AI practices with broader sustainability goals. This study emphasizes a strong connection between RAI and ESG, while also identifying significant gaps in public disclosures and internal governance. To address these challenges, we recommend that companies enhance board-level expertise in RAI, increase transparency in RAI policies, improve employee awareness and engagement in RAI, and set clear AI targets. These recommendations align with key ESG topics such as board and management oversight, policy development, disclosure and reporting, diversity, equity, inclusion, human rights, and labor management \cite{lee2024integrating}, thereby securing better ESG performance.
Implementing these recommendations allows companies to more effectively integrate RAI strategies that support both ethical AI development and long-term ESG objectives, ultimately leading to more responsible and transparent AI practices.

This study has some limitations that provide opportunities for future improvement. Although we included global companies in the interviews, 60\% of the engaged companies are based in Australia. This concentration may pose a challenge in generalizing the insights across different regions. In the next steps, we plan to include a more diverse set of companies from various geographical locations to enhance the representativeness of our findings.
Additionally, we relied solely on the interview responses without corroborating them with real-world practices, which means the study lacks direct evidence. To improve the validity of our conclusions, we intend to gather real-world data by conducting in-depth analyses of the companies' RAI practices in future research.
\vspace*{-8pt}

\bibliographystyle{IEEEtran}
\bibliography{bibliography}

\end{document}